\documentclass[a4paper,12pt]{article}

\usepackage{graphicx}
\usepackage{amsmath}
\usepackage{amssymb}
\usepackage{hyperref}
\usepackage{tabularx}
\newcolumntype{M}{>{$}c<{$}}
\usepackage[matrix,arrow,curve]{xy}

\numberwithin{equation}{section} \numberwithin{figure}{section}
\numberwithin{table}{section}

\def\papertitlepage{\baselineskip 3.5ex\thispagestyle{empty}}
\def\Title#1{\baselineskip 1cm \vspace{1.5cm}%
  \begin{center}{\Large\bf #1}\end{center}\vspace{0.5cm}}
\def\Authors#1{\begin{center}\renewcommand{\thefootnote}{\fnsymbol{footnote}}{\it #1}\end{center}}
\def\Abstract{\vspace{1.0cm}%
  \begin{center}{\large\bf Abstract}\end{center}}

\renewenvironment{thebibliography}{\pagebreak[3]\par\vspace{0.6em}
\begin{flushleft}{\large \bf References}\end{flushleft}
\vspace{-1.0em}

\begin{enumerate}\if@twocolumn\baselineskip=0.6em\itemsep -0.2em
\else\itemsep -0.2em\fi\labelsep 0.1em}{\end{enumerate} }

\DeclareMathDelimiter{\lcolon}{\mathopen}{operators}{"3A}{largesymbols}{"3A}
\DeclareMathDelimiter{\rcolon}{\mathclose}{operators}{"3A}{largesymbols}{"3A}

\def\+{\!\!+\!\!}

\def\dynkin(#1){(#1)}

\def\bra<#1|{\langle#1|}
\def\ket|#1>{|#1\rangle}
\def\braket<#1|#2>{\langle#1|#2\rangle}
\def\llangle{\langle\!\langle}
\def\rrangle{\rangle\!\rangle}
\def\bbra<#1|{\llangle#1|}
\def\kket|#1>{|#1\rrangle}
\def\bbraket<#1|#2>{\llangle#1|#2\rrangle}

\begin{document}
{\papertitlepage \vspace*{0cm} {\hfill
\begin{minipage}{4.2cm}
IF-USP 2011\par\noindent November, 2011
\end{minipage}}
\Title{Level truncation analysis of regularized identity based
solutions}
\Authors{{\sc E.~Aldo~Arroyo${}$\footnote{\tt
aldohep@fma.if.usp.br}}
\\
Instituto de F\'{i}sica, Universidade de S\~{a}o Paulo \\[-2ex]
C.P. 66.318 CEP 05315-970, S\~{a}o Paulo, SP, Brasil ${}$ }
} 

\vskip-\baselineskip
{\baselineskip .5cm \Abstract We evaluate the vacuum energy of
regularized identity based solutions through level truncation
computations in open bosonic string field theory. We show that the
level truncated solutions bring a value of the vacuum energy
expected for the tachyon vacuum in agreement with Sen's first
conjecture.

 }
\newpage
\setcounter{footnote}{0}
\tableofcontents

\section{Introduction}
In a previous work \cite{Arroyo:2010sy}, we have studied the
validity of a proposal for regularizing identity based solutions
\cite{Zeze:2010sr} in open string field theory
\cite{Witten:1985cc}. The regularization is based on the
realization that a one-parameter family of solutions, of the open
string field equations of motion, can be constructed by a gauge
transformation on an identity based solution
\cite{Arroyo:2010sy,Zeze:2010sr}
\begin{eqnarray}
\label{11} \Psi_\lambda = \mathcal{U}_\lambda (Q_B + \Psi_I)
\mathcal{U}_\lambda^{-1},
\end{eqnarray}
where $\Psi_I$ denotes the identity based solution and
$\mathcal{U}_\lambda$ is an element of the gauge transformation
defined as
\begin{eqnarray}
\label{12} \mathcal{U}_\lambda = 1+\lambda cBK, \;\;\;\;\;
\mathcal{U}_\lambda^{-1} = 1-\lambda cB \frac{K}{1+\lambda K}.
\end{eqnarray}

Explicitly the one-parameter family of solutions $\Psi_\lambda$,
which will be referred as the regularized solution, is obtained by
performing the gauge transformation on the identity based solution
$\Psi_I=c(1-K)$
\begin{eqnarray}
\label{psiintro1} \Psi_\lambda = c(1+\lambda K) Bc
\frac{1+(\lambda-1)K}{1+\lambda K} \,.
\end{eqnarray}
This solution interpolates between the identity based solution
$\Psi_I=\Psi_{\lambda\rightarrow 0}$ and Erler-Schnabl's solution
$\Psi_{E-S}=\Psi_{\lambda\rightarrow 1}$
\cite{Arroyo:2010fq,Erler:2009uj}. The value of the vacuum energy,
using the regularized solution, was computed by evaluating formal
CFT correlation functions in the so-called $KBc$ subalgebra
\cite{Zeze:2011zz} and by using $\mathcal{L}_0$ level truncation
computations \cite{Arroyo:2010sy}.

Nevertheless, it remains the analysis of the regularized solution
using the traditional Virasoro $L_0$ level truncation scheme
\cite{Kostelecky:1988ta,Kostelecky:1989nt,Moeller:2000xv,
Taylor:2002fy,Gaiotto:2002wy,Kishimoto:2011zz}. This analysis is
important since we want to know if the solution behaves as a
regular element in the state space constructed out of Fock states.
Specifically the analysis of the coefficients appearing in the
$L_0$ level expansion provides one criterion for the solution
being well defined
\cite{Murata:2011ex,Schnabl:2010tb,Takahashi:2007du}. Furthermore
the $L_0$ level expansion of the solution brings an additional way
to numerically test Sen's first conjecture.

To expand the regularized solution in the Virasoro basis of $L_0$
eigenstates, it is convenient to write an integral definition for
the rational function
\begin{eqnarray}
\frac{1+(\lambda-1)K}{1+\lambda K} = \int_0^{\infty}dt\,
e^{-\frac{t}{\lambda
}}\big[\frac{1}{\lambda}-\frac{(\lambda-1)}{\lambda}\partial_t\big]e^{-K
t} \,,
\end{eqnarray}
where $\partial_t \equiv \frac{\partial}{\partial t}$ and $e^{-Kt}
\equiv \Omega^t$ is the wedge state with $t\geq0$
\cite{Schnabl:2002gg,Schnabl:2005gv,Kiermaier:2007ba,Erler:2011tc,Bonora:2011ru,Bonora:2011ri}.
Let us mention that the wedge state can be expressed in terms of
the well known scaling operator $U_{r}$
\cite{Schnabl:2005gv,Okawa:2006vm,Fuchs:2006hw,Arroyo:2009ec,Erler:2006hw,Erler:2006ww}
\begin{eqnarray}
\Omega^t = U^{\dagger}_{t+1}  U_{t+1} | 0 \rangle,
\;\;\;\;\;\text{where}\;\;\;\;\;  U_{r} \equiv \Big(\frac{2}{r}
\Big)^{\mathcal{L}_0}.
\end{eqnarray}
The operator $\mathcal{L}_0$ is defined in the sliver frame
\cite{Erler:2009uj,Schnabl:2005gv,Arroyo:2011zt,AldoArroyo:2009hf}
\footnote{Remember that a point in the upper half plane $z$ is
mapped to a point in the sliver frame $\tilde z$ via the conformal
mapping $\tilde z= \frac{2}{\pi}\arctan z $. Note that we are
using the convention of \cite{Erler:2009uj} for the conformal
mapping.}, and it is related to the worldsheet energy-momentum
tensor as follows
\begin{eqnarray}
\mathcal{L}_0 = \oint \frac{d z}{2 \pi i} (1+z^{2}) (\arctan z) \,
T(z)= L_0 + \sum_{k=1}^{\infty} \frac{2(-1)^{k+1}}{4k^2-1} L_{2k}
\, .
\end{eqnarray}

Employing the above considerations and noting that under the
conformal map defined by $\tilde z=\frac{2}{\pi} \arctan z$ the
$c$ ghost transforms as $\tilde c(\tilde z)=\frac{2}{\pi}
\cos^{2}(\frac{\pi}{2} \tilde z) c \big(\tan (\frac{\pi}{2} \tilde
z)\big) $, we will show that the regularized solution can be
written like
\begin{align}
\label{psiintro2} \Psi_\lambda =&\frac{2}{\pi} \int_0^{\infty}dt\,
e^{-\frac{t}{\lambda }}
\big(\frac{1}{\lambda}-\frac{(\lambda-1)}{\lambda}\partial_t\big)
\Big[\big(\frac{t+1}{2}\big) \cos^2\big(\frac{\pi  t}{2 t+2}\big)
U^{\dag}_{t+1} c \Big( \tan \big(\frac{\pi t}{2 t+2}\big) \Big)
\Big]| 0 \rangle \nonumber
\\&+Q_B \Big\{ \frac{\lambda}{\pi}\int_0^{\infty}dt\,
e^{-\frac{t}{\lambda }}
\big(\frac{1}{\lambda}-\frac{(\lambda-1)}{\lambda}\partial_t\big)
\Big[\cos^2\big(\frac{\pi  t}{2 t+2}\big) U^{\dag}_{t+1}
\mathcal{B}^{\dag}_0 c \Big( \tan \big(\frac{\pi t}{2 t+2}\big)
\Big) \Big]| 0 \rangle \Big\}.
\end{align}
This expression allows us to expand the regularized solution using
states in the Virasoro basis of $L_0$ eigenstates. For example,
let us expand the regularized solution up to level two states
\begin{eqnarray}
\Psi_\lambda = t(\lambda) c_1 | 0 \rangle + u(\lambda) c_0 | 0
\rangle + v(\lambda) c_{-1} | 0 \rangle + w(\lambda) L_{-2}c_1 | 0
\rangle + \cdots + (Q_B\text{-exact terms}),
\end{eqnarray}
where the coefficients of the expansion $t(\lambda)$,
$u(\lambda)$, $v(\lambda)$ and $w(\lambda)$  are given by the
following integrals
\begin{align}
t(\lambda) &= \frac{2}{\pi} \int_0^{\infty}dt\,
e^{-\frac{t}{\lambda }}
\Big(\frac{1}{\lambda}-\frac{(\lambda-1)}{\lambda}\partial_t\Big)
\Big[ \frac{1}{4} (t+1)^2 \cos
^2\left(\frac{\pi  t}{2 t+2}\right) \Big], \\
u(\lambda) &= \frac{2}{\pi} \int_0^{\infty}dt\,
e^{-\frac{t}{\lambda }}
\Big(\frac{1}{\lambda}-\frac{(\lambda-1)}{\lambda}\partial_t\Big)
\Big[ \frac{1}{4} (t+1) \sin
\left(\frac{\pi  t}{t+1}\right) \Big], \\
v(\lambda) &= \frac{2}{\pi} \int_0^{\infty}dt\,
e^{-\frac{t}{\lambda }}
\Big(\frac{1}{\lambda}-\frac{(\lambda-1)}{\lambda}\partial_t\Big)
\Big[ \sin ^2\left(\frac{\pi  t}{2
t+2}\right) \Big], \\
w(\lambda) &= \frac{2}{\pi} \int_0^{\infty}dt\,
e^{-\frac{t}{\lambda }}
\Big(\frac{1}{\lambda}-\frac{(\lambda-1)}{\lambda}\partial_t\Big)
\Big[ \frac{1}{12} \left(3-t^2-2 t\right) \cos ^2\left(\frac{\pi
t}{2 t+2}\right) \Big].
\end{align}
These integrals are convergent provided that the parameter
$\lambda$ belongs to the interval $(0,+\infty)$. The range of
validity for this parameter $\lambda$ was also studied using the
$\mathcal{L}_0$ level expansion of the regularized solution
\cite{Arroyo:2010sy}. Let us comment that this result, relating to
the parameter $\lambda$, is quite similar to the well known case
of Schnabl-Okawa's one-parameter family of solutions analyzed in
\cite{Takahashi:2007du,Schnabl:2005gv,Okawa:2006vm,Fuchs:2006hw}
where it was shown that the parameter $\lambda$ belongs to the
interval $(0,1)$ and the limit case $\lambda \rightarrow 1$
precisely corresponds to the original Schnabl's analytic solution
\cite{Schnabl:2005gv}.

In this paper, by considering the traditional Virasoro $L_0$ level
expansion of the regularized solution $\Psi_\lambda$, we evaluate
the value of the vacuum energy for various values of the parameter
$\lambda
> 0$, and show that in the limit case
$\lambda \rightarrow 0$, which corresponds to the identity based
solution, the regularized solution brings a value of the vacuum
energy expected for the tachyon vacuum in agreement with Sen's
first conjecture \cite{Sen:1999mh,Sen:1999nx}. The results of our
computations provide an additional support to the fact that to
extract a finite value for the vacuum energy, the identity based
solution can be defined as the limit of a gauge equivalent
one-parameter family of solutions. Intuitively this approach works
since the vacuum energy is a gauge invariant quantity so that its
value is insensitive to the change of the parameter, thus after
performing the computations we can safely take the limit.
Nevertheless the computation of the relevant gauge invariant
quantities using directly the identity based solution remains as
an unsolved problem. As noted in \cite{Schnabl:2010tb}, in order
to solve this issue, it should be interesting to rephrase the
problem in terms of distribution theory \cite{Bonora:2011ns}. We
hope that explorations in these directions also might give a
lesson to define the space of string fields appropriately.

In addition to the level truncation analysis of the regularized
solution $\Psi_\lambda$, we will study the level truncation of
another solution $\hat \Psi_\lambda$ which satisfies the string
field reality condition \footnote{In open string field theory, the
string field reality condition over a string field $\Phi$ is given
by $\Phi^{\ddagger} = \Phi$, where the operator $\ddagger$ is
defined by the composition of BPZ and Hermitian conjugation
\cite{Gaberdiel:1997ia}.}
\begin{eqnarray}
\label{introx01} \hat \Psi_\lambda =
\sqrt{\frac{1+(\lambda-1)K}{1+\lambda K}} (c+\lambda cKBc)
\sqrt{\frac{1+(\lambda-1)K}{1+\lambda K}}.
\end{eqnarray}
Because of the square root in (\ref{introx01}), this real solution
$\hat \Psi_\lambda$ seems to be more complicated than the previous
one $\Psi_\lambda$ \footnote{In what follows, the regularized
solution $\Psi_\lambda$ will be referred as the non-real
solution.}. But for some purposes the real solution is more
convenient. For instance, at some fixed level the number of terms
contained in the level expansion of the real solution will be less
than the one given in the non-real solution. This difference in
the number of terms will be reflected in the evaluation of the
vacuum energy, at each level using the real solution we will get
less terms to be computed than the ones obtained from the non-real
solution. Another advantage of the real solution is related to the
convergence property regarding to the evaluation of the vacuum
energy. Each time we increase the level of the truncated solution,
the value of the vacuum energy gets closer to the expected value
predicted from Sen's first conjecture. For the case of the real
solution it turns out that the convergence to this expected value
is faster than the one using the non-real solution.

This paper is organized as follows. In section 2, we study the way
for regularizing an identity based solution in open bosonic string
field theory and by expanding the regularized solution in the
Virasoro basis of $L_0$ eigenstates, we compute the vacuum energy.
In section 3, we introduce another one-parameter family of
solutions which satisfies the string field reality condition. By
performing a gauge transformation, we show that this real solution
can be related to the non-real solution. The level truncation
computation of the vacuum energy is analyzed in a similar manner
as for the non-real solution. In section 4, a summary and further
directions of exploration are given.

\section{Regularized identity based solution in open string field theory}
In this section, first we are going to review the way for
regularizing an identity based solution in open bosonic string
field theory and then we will analyze the solution from the level
truncation point of view.

\subsection{Derivation of the regularized solution}
Let us mention that the main results showed in this subsection
will be extracted from the references
\cite{Arroyo:2010sy,Zeze:2010sr,Arroyo:2010fq}. As discussed in
\cite{Arroyo:2010fq} using the methods of
\cite{Erler:2006hw,Erler:2006ww}, an identity based solution in
open bosonic string field theory is given by
\begin{eqnarray}
\label{identity01} \Psi_I=c(1-K), \,
\end{eqnarray}
where the basic string fields $c$ and $K$ (together with $B$) can
be written, using the operator representation
\cite{Schnabl:2005gv}, as follows
\begin{eqnarray}
\label{K01} K &\rightarrow& \frac{1}{2} \hat{\mathcal{L}}
U_{1}^\dag U_{1} |0\rangle \, ,
\\
\label{B01} B &\rightarrow& \frac{1}{2} \hat{\mathcal{B}}
U_{1}^\dag U_{1} |0\rangle \, ,
\\
\label{c01} c &\rightarrow&   U_{1}^\dag U_{1} \tilde c
(0)|0\rangle \, .
\end{eqnarray}

The operators $\hat{\mathcal{L}}$, $\hat{\mathcal{B}}$ and $\tilde
c(0)$ are defined in the sliver frame \cite{Erler:2009uj}
\footnote{Remember that a point in the upper half plane $z$ is
mapped to a point in the sliver frame $\tilde z$ via the conformal
mapping $\tilde z= \frac{2}{\pi}\arctan z $. Note that we are
using the convention of \cite{Erler:2009uj} for the conformal
mapping.}, and they are related to the worldsheet energy-momentum
tensor, the $b$ and $c$ ghosts fields respectively, for instance
\begin{eqnarray}
\label{Lhat01} \hat{\mathcal{L}} &\equiv& \mathcal{L}_{0} +
\mathcal{L}^{\dag}_0 = \oint \frac{d z}{2 \pi i} (1+z^{2})
(\arctan z+\text{arccot} z) \,
T(z) \, , \\
\label{Bhat01} \hat{\mathcal{B}} &\equiv& \mathcal{B}_{0} +
\mathcal{B}^{\dag}_0 = \oint \frac{d z}{2 \pi i} (1+z^{2})
(\arctan z+\text{arccot} z) \, b(z) \, ,
\end{eqnarray}
while the operator $U_{1}^\dag U_{1}$ in general is given by
$U^\dag_r U_r = e^{\frac{2-r}{2} \hat{\mathcal{L}}}$, so we have
chosen $r=1$, note that the string field $U_{1}^\dag U_{1}
|0\rangle$ represents to the identity string field $1 \rightarrow
U_{1}^\dag U_{1} |0\rangle$
\cite{Schnabl:2005gv,Okawa:2006vm,Erler:2006hw,Erler:2006ww}.

Using the operator representation (\ref{K01})-(\ref{c01}) of the
string fields $K$, $B$ and $c$, we can show that these fields
satisfy the algebraic relations
\begin{eqnarray}
\label{eq2} \{B,c\}=1\, , \;\;\;\;\;\;\; [B,K]=0 \, ,
\;\;\;\;\;\;\; B^2=c^2=0 \, ,
\end{eqnarray}
and have the following BRST variations
\begin{eqnarray}
\label{eq3} Q_BK=0 \, , \;\;\;\;\;\; Q_BB=K \, , \;\;\;\;\;\;
Q_Bc=cKc \, .
\end{eqnarray}

As it is shown in \cite{Arroyo:2010fq} the direct evaluation of
the vacuum energy using the identity based solution
(\ref{identity01}) brings ambiguous result. This phenomenon, as it
was noted in \cite{Zeze:2010sr}, is due to the fact that a naive
evaluation of the classical action in terms of CFT methods tends
to be indefinite since it corresponds to a correlator on vanishing
strip. Recently this problem was overcome and a proposal for
regularizing the identity based solution (\ref{identity01}) has
been developed in \cite{Zeze:2010sr}.

The regularized solution $\Psi_\lambda$ is obtained by considering
one-parameter family of classical solutions
\begin{eqnarray}
\label{regularsol1} \Psi_\lambda = \mathcal{U}_\lambda Q_B
\mathcal{U}_\lambda^{-1} + \mathcal{U}_\lambda \Psi_I
\mathcal{U}_\lambda^{-1} \, ,
\end{eqnarray}
where $\Psi_I$ is the identity based solution (\ref{identity01})
and
\begin{eqnarray}
\label{gauge01} \mathcal{U}_\lambda=1+\lambda cBK\, , \;\;\;
\mathcal{U}_\lambda^{-1}=1-\lambda cBK \frac{1}{1+\lambda K}
\end{eqnarray}
is an element of the gauge transformation
\cite{Zeze:2010sr,Arroyo:2010fq}. Using (\ref{identity01}),
(\ref{regularsol1}) and (\ref{gauge01}), it is almost easy to
derive the following regularized solution
\begin{eqnarray}
\label{regularsol2} \Psi_\lambda = c(1+\lambda K) Bc
\frac{1+(\lambda-1)K}{1+\lambda K} \,.
\end{eqnarray}

In order to simplify the evaluation of the vacuum energy, it is
convenient to write the regularized solution $\Psi_\lambda$ as an
expression containing an exact BRST term
\begin{eqnarray}
\label{exactbrst1} \Psi_\lambda = c
\frac{1+(\lambda-1)K}{1+\lambda K}+ Q_B \Big\{ \lambda Bc
\frac{1+(\lambda-1)K}{1+\lambda K} \Big\} \,.
\end{eqnarray}
Note that this regularized solution interpolates between the
identity based solution (\ref{identity01}) which corresponds to
the case $\lambda\rightarrow 0$, and the Erler-Schnabl's solution
\cite{Erler:2009uj} which corresponds to the case
$\lambda\rightarrow 1$.

\subsection{Level expansion of the regularized solution}
To expand the previous regularized solution in the Virasoro basis
of $L_0$ eigenstates, it is convenient to write an integral
definition for the rational function
\begin{eqnarray}
\label{rational01xx} \frac{1+(\lambda-1)K}{1+\lambda K} =
\int_0^{\infty}dt\, e^{-\frac{t}{\lambda
}}\big[\frac{1}{\lambda}-\frac{(\lambda-1)}{\lambda}\partial_t\big]e^{-K
t} \,,
\end{eqnarray}
where $\partial_t \equiv \frac{\partial}{\partial t}$ and $e^{-Kt}
\equiv \Omega^t$ is the wedge state with $t\geq0$. Let us mention
that the wedge state can be expressed in terms of the well known
scaling operator $U_{r}$
\cite{Schnabl:2005gv,Okawa:2006vm,Fuchs:2006hw,Arroyo:2009ec,Erler:2006hw,Erler:2006ww}
\begin{eqnarray}
\label{omeome1} \Omega^t = U^{\dagger}_{t+1}  U_{t+1}| 0 \rangle,
\;\;\;\;\;\text{where}\;\;\;\;\;  U_{r} \equiv \Big(\frac{2}{r}
\Big)^{\mathcal{L}_0}.
\end{eqnarray}

For level truncation computations, it is useful to write the
scaling operator as the following canonical ordered form
\begin{eqnarray}
\label{uu1} U_r = \Big(\frac{2}{r}\Big)^{L_0} e^{u_{2,r} L_2}
e^{u_{4,r} L_4} e^{u_{6,r} L_6} e^{u_{8,r} L_8}e^{u_{10,r} L_{10}}
 \cdots,
\end{eqnarray}
where the operators $L_n$ are the usual Virasoro generators. To
find the coefficients $u_{n,r}$ appearing in the exponentials, we
use
\begin{align}
\frac{r}{2} \tan (\frac{2}{r} \arctan z) &= \lim_{N \rightarrow
\infty} \big[f_{2,u_{2,r}} \circ  f_{4,u_{4,r}} \circ
f_{6,u_{6,r}} \circ f_{8,u_{8,r}} \circ
f_{10,u_{10,r}} \circ \cdots \circ f_{N,u_{N,r}}(z)\big] \nonumber \\
&= \lim_{N \rightarrow \infty}\big[ f_{2,u_{2,r}} ( f_{4,u_{4,r}}
( f_{6,u_{6,r}} ( f_{8,u_{8,r}} ( f_{10,u_{10,r}}(\cdots
(f_{N,u_{N,r}}(z)) \dots )))))  \big],
\end{align}
where
\begin{eqnarray}
f_{n,u_{n,r}}(z) = \frac{z}{(1-u_{n,r} n z^n)^{1/n}}.
\end{eqnarray}
Using the above expressions, we get
\begin{align}
\label{tt2}
u_{2,r} &= \frac{4-r^2}{3 r^2}, \\
u_{4,r} &= \frac{r^4-16}{30 r^4}, \\
u_{6,r} &= -\frac{16 \left(r^6-21 r^2+20\right)}{945 r^6}, \\
u_{8,r} &=\frac{109 r^8-2688 r^4+5120 r^2-5376}{11340 r^8},\\
\label{tt10} u_{10,r} &= -\frac{16 \left(9 r^{10}-253 r^6+660
r^4-1056
   r^2+640\right)}{22275 r^{10}}.
\end{align}

Plugging the definition of the rational function
(\ref{rational01xx}) into the expression for the regularized
solution (\ref{exactbrst1}), we obtain
\begin{eqnarray}
\label{exactbrst1level01} \Psi_\lambda = \int_0^{\infty}dt\,
e^{-\frac{t}{\lambda
}}\big[\frac{1}{\lambda}-\frac{(\lambda-1)}{\lambda}\partial_t\big]
c \Omega^t+ Q_B \Big\{ \lambda \int_0^{\infty}dt\,
e^{-\frac{t}{\lambda
}}\big[\frac{1}{\lambda}-\frac{(\lambda-1)}{\lambda}\partial_t\big]
Bc \Omega^t \Big\} \,. \;\;\;\;
\end{eqnarray}
Using the operator representation for the $c$ ghost (\ref{c01}),
the expression for the wedge state (\ref{omeome1}), and noting
that under the conformal map $\tilde z=\frac{2}{\pi} \arctan z$,
the $c$ ghost transforms as $\tilde c(\tilde z)=\frac{2}{\pi}
\cos^{2}(\frac{\pi}{2} \tilde z) c \big(\tan (\frac{\pi}{2} \tilde
z)\big)$, from equation (\ref{exactbrst1level01}) we derive
\begin{align}
\label{psilevelxx2} \Psi_\lambda =&\frac{2}{\pi}
\int_0^{\infty}dt\, e^{-\frac{t}{\lambda }}
\big(\frac{1}{\lambda}-\frac{(\lambda-1)}{\lambda}\partial_t\big)
\Big[\big(\frac{t+1}{2}\big) \cos^2\big(\frac{\pi  t}{2 t+2}\big)
U^{\dag}_{t+1} c \Big( \tan \big(\frac{\pi t}{2 t+2}\big) \Big)
\Big]| 0 \rangle \nonumber
\\&+Q_B \Big\{ \frac{\lambda}{\pi}\int_0^{\infty}dt\,
e^{-\frac{t}{\lambda }}
\big(\frac{1}{\lambda}-\frac{(\lambda-1)}{\lambda}\partial_t\big)
\Big[\cos^2\big(\frac{\pi  t}{2 t+2}\big) U^{\dag}_{t+1}
\mathcal{B}^{\dag}_0 c \Big( \tan \big(\frac{\pi t}{2 t+2}\big)
\Big) \Big]| 0 \rangle \Big\}.
\end{align}

Since in the next subsection it will be argued that to evaluate
the vacuum energy, we only need to consider the first term of the
regularized solution (\ref{psilevelxx2}), let us study in some
detail this first term
\begin{align}
\label{psifirst1} \Psi^{(1)}_\lambda \equiv \frac{2}{\pi}
\int_0^{\infty}dt\, e^{-\frac{t}{\lambda }}
\big(\frac{1}{\lambda}-\frac{(\lambda-1)}{\lambda}\partial_t\big)
\Big[\big(\frac{t+1}{2}\big) \cos^2\big(\frac{\pi  t}{2 t+2}\big)
U^{\dag}_{t+1} c \Big( \tan \big(\frac{\pi t}{2 t+2}\big) \Big)
\Big]| 0 \rangle. \;\;
\end{align}
Using the expression (\ref{uu1}) for the operator $U_r$, from
equation (\ref{psifirst1}) we obtain an expression for
$\Psi^{(1)}_\lambda$ which is ready to be used in level truncation
computations
\begin{align}
\label{psifirst1level} \Psi^{(1)}_\lambda = \frac{2}{\pi}
\int_0^{\infty} dt\,e^{-\frac{t}{\lambda }}
\big(\frac{1}{\lambda}-\frac{(\lambda-1)}{\lambda}\partial_t\big)
\Big[\big(\frac{t+1}{2}\big)^2 \cos^2\big(\frac{\pi  t}{2
t+2}\big) \widetilde{U}_{t+1} c \Big( \frac{2}{t+1} \tan
\big(\frac{\pi t}{2 t+2}\big) \Big) \Big]| 0 \rangle, \;\;\;\;\;
\end{align}
where
\begin{eqnarray}
\label{uux2} \widetilde{U}_{t+1} \equiv  \cdots e^{u_{10,t+1}
L_{-10}} e^{u_{8,t+1} L_{-8}} e^{u_{6,t+1} L_{-6}} e^{u_{4,t+1}
L_{-4}}e^{u_{2,t+1} L_{-2}}.
\end{eqnarray}
The coefficients $u_{n,t+1}$ appearing in the exponentials can be
computed by performing the substitution $r \rightarrow t+1$ in the
set of equations (\ref{tt2})-(\ref{tt10}).

By writing the $c$ ghost in terms of its modes $c(z)=\sum_{m}
c_m/z^{m-1}$ and employing equations (\ref{psifirst1level}) and
(\ref{uux2}), we can expand $\Psi^{(1)}_\lambda$ in terms of
elements contained in the Virasoro basis of $L_0$ eigenstates. For
example, let us expand $\Psi^{(1)}_\lambda$ up to level two states
\begin{eqnarray}
\Psi^{(1)}_\lambda= t(\lambda) c_1 | 0 \rangle + u(\lambda) c_0 |
0 \rangle + v(\lambda) c_{-1} | 0 \rangle + w(\lambda) L_{-2}c_1 |
0 \rangle + \cdots ,
\end{eqnarray}
where the coefficients of the expansion $t(\lambda)$,
$u(\lambda)$, $v(\lambda)$ and $w(\lambda)$  are given by the
following integrals
\begin{align}
t(\lambda) &= \frac{2}{\pi} \int_0^{\infty}dt\,
e^{-\frac{t}{\lambda }}
\Big(\frac{1}{\lambda}-\frac{(\lambda-1)}{\lambda}\partial_t\Big)
\Big[ \frac{1}{4} (t+1)^2 \cos
^2\left(\frac{\pi  t}{2 t+2}\right) \Big], \\
u(\lambda) &= \frac{2}{\pi} \int_0^{\infty}dt\,
e^{-\frac{t}{\lambda }}
\Big(\frac{1}{\lambda}-\frac{(\lambda-1)}{\lambda}\partial_t\Big)
\Big[ \frac{1}{4} (t+1) \sin
\left(\frac{\pi  t}{t+1}\right) \Big], \\
v(\lambda) &= \frac{2}{\pi} \int_0^{\infty}dt\,
e^{-\frac{t}{\lambda }}
\Big(\frac{1}{\lambda}-\frac{(\lambda-1)}{\lambda}\partial_t\Big)
\Big[ \sin ^2\left(\frac{\pi  t}{2
t+2}\right) \Big], \\
w(\lambda) &= \frac{2}{\pi} \int_0^{\infty}dt\,
e^{-\frac{t}{\lambda }}
\Big(\frac{1}{\lambda}-\frac{(\lambda-1)}{\lambda}\partial_t\Big)
\Big[ \frac{1}{12} \left(3-t^2-2 t\right) \cos ^2\left(\frac{\pi
t}{2 t+2}\right) \Big].
\end{align}
These integrals are convergent provided that the parameter
$\lambda$ belongs to the interval $(0,+\infty)$. The range of
validity for this parameter $\lambda$ was also studied using the
$\mathcal{L}_0$ level expansion of the regularized solution
\cite{Arroyo:2010sy}. Let us comment that this result, relating to
the parameter $\lambda$, is quite similar to the well known case
of Schnabl-Okawa's one-parameter family of solutions analyzed in
\cite{Takahashi:2007du,Schnabl:2005gv,Okawa:2006vm,Fuchs:2006hw}
where it was shown that the parameter $\lambda$ belongs to the
interval $(0,1)$ and the limit case $\lambda \rightarrow 1$
precisely corresponds to the original Schnabl's analytic solution
\cite{Schnabl:2005gv}.

\subsection{Evaluation of the vacuum energy}
Assuming the validity of the string field equation of motion when
contracted with the solution itself \cite{Arroyo:2010sy}, we can
show that the normalized value of the vacuum energy is given by
\begin{eqnarray}
\label{normaener1} E = \frac{\pi^2}{3} \langle \Psi_\lambda, Q_B
\Psi_\lambda\rangle.
\end{eqnarray}
Since the regularized solution $\Psi_\lambda$ can be written as an
expression containing an exact BRST term, to compute the
normalized value of the vacuum energy (\ref{normaener1}), we only
need to consider the first term of (\ref{psilevelxx2})
\begin{eqnarray}
\label{normaener2} E = \frac{\pi^2}{3} \langle \Psi^{(1)}_\lambda,
Q_B \Psi^{(1)}_\lambda\rangle.
\end{eqnarray}

As described in
\cite{Erler:2009uj,Takahashi:2007du,Schnabl:2005gv,Arroyo:2009ec},
it is convenient to replace the string field $\Psi^{(1)}_\lambda$
with $z^{L_0} \Psi^{(1)}_\lambda$ in the $L_0$ level truncation
scheme, so that states in the $L_0$ level expansion of the
solution acquire different integer powers of $z$ at different
levels. As we are going to see, the parameter $z$ is needed
because we will need to express the normalized value of the vacuum
energy (\ref{normaener2}) as a formal power series expansion if we
want to use Pad\'{e} approximants
\cite{Arroyo:2009ec,AldoArroyo:2008zm}. After doing our
calculations, we will simply set $z = 1$.

Using (\ref{psifirst1level}) and (\ref{uux2}) the string field
$\Psi^{(1)}_\lambda$ can be readily expanded and the individual
coefficients can be numerically integrated. For instance,
employing some numerical values for the parameter $\lambda$, we
obtain
\begin{align}
\label{psi1over10} \Psi^{(1)}_{\lambda=1/10} &=& 0.43430476\, c_1|
0 \rangle+0.47429914\,c_0| 0 \rangle+0.18754065\, c_{-1}| 0
\rangle+0.00492478\, L_{-2}c_1| 0 \rangle \nonumber \\
&&+0.09189138\, c_{-2}| 0 \rangle+0.30188401\, L_{-2}c_0| 0
\rangle + 0.05141509\, c_{-3}| 0 \rangle \;\;\;\;\;\;\;\;\;\;\;\;\;\;\;\;\;\;\;\;\;\;\;\;\;\; \nonumber \\
&& +0.07875502\, L_{-4}c_1| 0 \rangle+0.10209574\,L_{-2}c_{-1}| 0
\rangle-0.13289997\,L_{-2}L_{-2}c_1| 0 \rangle+\cdots,\\
\label{psi2over10} \Psi^{(1)}_{\lambda=2/10} &=& 0.40070791\, c_1|
0 \rangle+0.43550897\,c_0| 0 \rangle+0.25391982\, c_{-1}| 0
\rangle-0.00600265\, L_{-2}c_1| 0 \rangle \nonumber \\
&&+0.17221906\, c_{-2}| 0 \rangle+0.19467134\, L_{-2}c_0| 0
\rangle +0.12798546\, c_{-3}| 0 \rangle \;\;\;\;\;\;\;\;\;\;\;\;\;\;\;\;\;\;\;\;\;\;\;\;\;\; \nonumber \\
&& +0.01431572\, L_{-4}c_1| 0 \rangle+0.08883011\,L_{-2}c_{-1}| 0
\rangle-0.09200582\,L_{-2}L_{-2}c_1| 0 \rangle+\cdots,\\
\label{psi3over10} \Psi^{(1)}_{\lambda=3/10} &=& 0.37467203\, c_1|
0 \rangle+0.39938623\,c_0| 0 \rangle+0.27811748\, c_{-1}| 0
\rangle-0.00538991\, L_{-2}c_1| 0 \rangle \nonumber \\
&&+0.21805409\, c_{-2}| 0 \rangle+0.13091926\, L_{-2}c_0| 0
\rangle +0.18356534\, c_{-3}| 0 \rangle \;\;\;\;\;\;\;\;\;\;\;\;\;\;\;\;\;\;\;\;\;\;\;\;\;\; \nonumber \\
&& +0.03887291\, L_{-4}c_1| 0 \rangle+0.06563034\,L_{-2}c_{-1}| 0
\rangle-0.06299155\,L_{-2}L_{-2}c_1| 0 \rangle+\cdots,\\
\label{psi4over10} \Psi^{(1)}_{\lambda=4/10} &=& 0.35396897\, c_1|
0 \rangle+0.36790312\,c_0| 0 \rangle+0.28461930\, c_{-1}| 0
\rangle-0.00065616\, L_{-2}c_1| 0 \rangle \nonumber \\
&&+0.24293724\, c_{-2}| 0 \rangle+0.08996485\, L_{-2}c_0| 0
\rangle + 0.21992140\, c_{-3}| 0 \rangle \;\;\;\;\;\;\;\;\;\;\;\;\;\;\;\;\;\;\;\;\;\;\;\;\;\; \nonumber \\
&& +0.02519440\, L_{-4}c_1| 0 \rangle +0.04537003\,L_{-2}c_{-1}| 0
\rangle-0.04177194\,L_{-2}L_{-2}c_1| 0 \rangle+\cdots,\\
\label{psi5over10} \Psi^{(1)}_{\lambda=5/10} &=& 0.33709802\, c_1|
0 \rangle+0.34076058\,c_0| 0 \rangle+0.28293698\, c_{-1}| 0
\rangle+0.00552825\, L_{-2}c_1| 0 \rangle \nonumber \\
&&+0.25561048\, c_{-2}| 0 \rangle+0.06216305\, L_{-2}c_0| 0
\rangle +0.24287102\, c_{-3}| 0 \rangle \;\;\;\;\;\;\;\;\;\;\;\;\;\;\;\;\;\;\;\;\;\;\;\;\;\; \nonumber \\
&& +0.01431572\, L_{-4}c_1| 0 \rangle+0.02931597\,L_{-2}c_{-1}| 0
\rangle-0.02570229\,L_{-2}L_{-2}c_1| 0 \rangle+\cdots.
\end{align}

Once we have the level expansion of the string field
$\Psi^{(1)}_\lambda$, we can compute the normalized value of the
vacuum energy. By performing the replacement $\Psi^{(1)}_\lambda
\rightarrow z^{L_0} \Psi^{(1)}_\lambda$ from equation
(\ref{normaener2}), we define
\begin{eqnarray}
\label{normaener1zz} E_\lambda(z) \equiv \frac{\pi^2}{3} \langle
z^{L_0} \Psi^{(1)}_\lambda, Q_B  z^{L_0}
\Psi^{(1)}_\lambda\rangle.
\end{eqnarray}
For example, plugging the level expansions
(\ref{psi1over10})-(\ref{psi5over10}) into the definition
(\ref{normaener1zz}), we obtain
\begin{eqnarray}
\label{normaener1zz1over10} E_{\lambda=1/10}(z)
=-\frac{0.620536}{z^2}-1.480175-0.097159 z^2+0.730101 z^4-0.680451
z^6+\cdots, \;\;\;\;\; \\
\label{normaener1zz2over10} E_{\lambda=2/10}(z)
=-\frac{0.528243}{z^2}-1.247965-0.241727 z^2+0.882372 z^4+0.064902
z^6+\cdots, \;\;\;\;\; \\
\label{normaener1zz3over10} E_{\lambda=3/10}(z)
=-\frac{0.461828}{z^2}-1.049529-0.283676 z^2+0.751339 z^4+0.313303
z^6+\cdots, \;\;\;\;\; \\
\label{normaener1zz4over10} E_{\lambda=4/10}(z)
=-\frac{0.412200}{z^2}-0.890585-0.270186 z^2+0.575221 z^4+0.327130
z^6+\cdots, \;\;\;\;\;\\
\label{normaener1zz5over10} E_{\lambda=5/10}(z)
=-\frac{0.373844}{z^2}-0.764024-0.232088 z^2+0.418196 z^4+0.244691
z^6+\cdots. \;\;\;\;\;
\end{eqnarray}

As a pedagogical illustration of the numerical method based on
Pad\'{e} approximants, let us compute in detail the normalized
value of the vacuum energy using the expansion
(\ref{normaener1zz5over10}) which corresponds to the case
$\lambda=1/2$. First, we express $E_{\lambda=1/2}(z)$ as the
rational function $P^4_{2+4}(\lambda,z)$
\begin{eqnarray}
\label{ss2} E_{\lambda=1/2}(z)=P^4_{2+4}(1/2,z)=\frac{1}{z^2}
\Big[\frac{a_0+a_1z+a_2z^2+a_3z^3+a_4z^4
}{1+b_1z+b_2z^2+b_3z^3+b_4z^4} \Big]\, .
\end{eqnarray}
Expanding the right hand side of (\ref{ss2}) around $z=0$ up to
sixth order in $z$ and equating the coefficients of $z^{-2}$,
$z^{-1}$, $z^{0}$, $z^{1}$, $z^{2}$, $z^{3}$, $z^{4}$, $z^{5}$,
$z^{6}$ with the expansion (\ref{normaener1zz5over10}), we get a
system of seven algebraic equations for the unknown coefficients
$a_0$, $a_1$, $a_2$, $a_3$, $a_4$, $b_1$, $b_2$, $b_3$ and $b_4$.
Solving those equations we get
\begin{eqnarray}
a_0 = -0.37384441, \;\;\; a_1=0, \;\;\; a_2=-0.67402070,
\;\;\; a_3=0, \;\;\; a_4=-0.28011517, \;\;\;\; \\
b_1 = 0, \;\;\; b_2=-0.24075156, \;\;\; b_3=0, \;\;\;
b_4=0.62049213. \;\;\;\;\;\;\;\;\;\;\;\;\;\;\;\;\;\;\;\;\;
\end{eqnarray}
Replacing the value of these coefficients into the definition of
$P^4_{2+4}(1/2,z)$ (\ref{ss2}), and evaluating this at $z=1$, we
get the following normalized value of the vacuum energy
\begin{eqnarray}
\label{vacum01x} P^4_{2+4}(1/2,z=1) = -0.96248549.
\end{eqnarray}

To evaluate higher Pad\'{e} approximants $P^n_{2+n}(\lambda,z)$,
we need to know the expansion of the normalized value of the
vacuum energy $E_\lambda(z)$ up to the order $z^{2n-2}$. We have
computed this $z$-dependent energy using the components of the
string field $\Psi^{(1)}_\lambda$ up to level $12$. The results of
our calculations are summarized in table \ref{resultsx1}. As we
can see, the normalized value of the vacuum energy computed
numerically using Pad\'{e} approximants confirm Sen's first
conjecture \cite{Sen:1999mh,Sen:1999nx}. Although in the case
where $\lambda \rightarrow 0$ and $\lambda > 1$, the convergence
to the expected answer gets quite slow, by considering higher
level contributions, we will eventually reach to the right value
of the vacuum energy $E \rightarrow -1$. The particular case
$\lambda = 1$ corresponds to the Erler-Schnabl's solution which
was already studied in \cite{Erler:2009uj}.

\begin{table}[ht]
\caption{The Pad\'{e} approximation for the normalized value of
the vacuum energy $\frac{\pi^2}{3} \langle z^{L_0}
\Psi^{(1)}_\lambda, Q_B z^{L_0} \Psi^{(1)}_\lambda\rangle$
evaluated at $z=1$. The results show the $P_{2+n}^{n}$ Pad\'{e}
approximation for various values of the parameter $\lambda$.}
\centering
\begin{tabular}{|c|c|c|c|c|c|}
\hline
  &  $P^{n}_{2+n}[\lambda = \frac{1}{10}]$   &  $P^{n}_{2+n}[\lambda = \frac{2}{10}]$ &
  $P^{n}_{2+n}[\lambda = \frac{3}{10} ]$ & $P^{n}_{2+n}[\lambda = \frac{4}{10} ]$ & $P^{n}_{2+n}[\lambda = \frac{5}{10}]$ \\
    \hline
$n=0$   &  $-0.62053698$ &  $-0.52824370$ & $-0.46182884$ & $-0.41220084$ & $-0.37384441$\\
\hline $n=4$   & $-1.87862775$ & $-1.43543509$ & $-1.19912518$ & $-1.05497540$ & $-0.96248549$ \\
\hline $n=8$  & $-1.38144216$ & $-1.26867724$ & $-1.12709179$ & $-1.03838541$ & $-0.99164341$ \\
\hline $n=12$ & $-1.37921868$ & $-1.23797300$ & $-1.11310799$ & $-1.03746194$ & $-0.85782931$ \\
\hline
\end{tabular}
\label{resultsx1}

\begin{tabular}{|c|c|c|c|c|c|}
\hline
  &  $P^{n}_{2+n}[\lambda = 1]$   &  $P^{n}_{2+n}[\lambda = \frac{12}{10}]$ &
  $P^{n}_{2+n}[\lambda = \frac{13}{10} ]$ & $P^{n}_{2+n}[\lambda = \frac{14}{10} ]$ & $P^{n}_{2+n}[\lambda = \frac{15}{10} ]$ \\
    \hline
$n=0$   &  $-0.26608479$ &  $-0.24232453$ & $-0.23260344$ & $-0.22399489$ & $-0.21631929$\\
\hline $n=4$   & $-0.67935543$ & $-0.52264016$ & $-0.46558572$ & $-0.41942773$ & $-0.38129318$ \\
\hline $n=8$  & $-0.93565531$ & $-0.90981927$ & $-0.89842271$ & $-0.88782546$ & $-0.87788963$ \\
\hline $n=12$ & $-0.94057422$ & $-0.90945338$ & $-0.89948567$ & $-0.88818475$ & $-0.87439427$ \\
\hline
\end{tabular}

\end{table}

It is interesting to observe that we can use another
identity-based solution for $\lambda=0$, for instance
$\Psi_0=\alpha c-cK$, with $\alpha$ being an arbitrary parameter.
Employing the gauge transformation (\ref{regularsol1}) and
(\ref{gauge01}) with $\Psi_I=c(\alpha -K)$, we can show that the
resulting two-parameter family of solutions is given by
\begin{eqnarray}
\label{regularsol2extra} \Psi_{\lambda,\alpha} = c(1+\lambda K) Bc
\frac{\alpha+(\alpha\lambda-1)K}{1+\lambda K} \,.
\end{eqnarray}
To simplify the evaluation of the vacuum energy for this solution,
it is convenient to write $\Psi_{\lambda,\alpha}$ as an expression
containing an exact BRST term
\begin{eqnarray}
\label{exactbrst1extra} \Psi_{\lambda,\alpha} = c
\frac{\alpha+(\alpha\lambda-1)K}{1+\lambda K}+ Q_B \Big\{ \lambda
Bc \frac{\alpha+(\alpha\lambda-1)K}{1+\lambda K} \Big\} \,.
\end{eqnarray}

Following the same procedures we used for the previous asymmetric
solution, we obtain an expression for the first term of the
right-hand side of (\ref{exactbrst1extra})
$\Psi^{(1)}_{\lambda,\alpha}$ which is ready for level truncation
computations,
\begin{align}
\label{psifirst1levelextra} \Psi^{(1)}_{\lambda,\alpha} =
\frac{2}{\pi} \int_0^{\infty} dt\,e^{-\frac{t}{\lambda }}
\big(\frac{\alpha}{\lambda}-\frac{(\alpha\lambda-1)}{\lambda}\partial_t\big)
\Big[\big(\frac{t+1}{2}\big)^2 \cos^2\big(\frac{\pi  t}{2
t+2}\big) \widetilde{U}_{t+1} c \Big( \frac{2}{t+1} \tan
\big(\frac{\pi t}{2 t+2}\big) \Big) \Big]| 0 \rangle. \;\;\;\;
\end{align}
For example, let us expand $\Psi^{(1)}_{\lambda,\alpha}$ up to
level two states,
\begin{eqnarray}
\label{psilaal}\Psi^{(1)}_{\lambda,\alpha}= t(\lambda,\alpha) c_1
| 0 \rangle + u(\lambda,\alpha) c_0 | 0 \rangle +
v(\lambda,\alpha) c_{-1} | 0 \rangle + w(\lambda,\alpha) L_{-2}c_1
| 0 \rangle + \cdots ,
\end{eqnarray}
where the coefficients of the expansion, $t(\lambda,\alpha)$,
$u(\lambda,\alpha)$, $v(\lambda,\alpha)$ and $w(\lambda,\alpha)$,
are given by the following integrals
\begin{align}
t(\lambda,\alpha) &= \frac{2}{\pi} \int_0^{\infty}dt\,
e^{-\frac{t}{\lambda }}
\Big(\frac{\alpha}{\lambda}-\frac{(\alpha\lambda-1)}{\lambda}\partial_t\Big)
\Big[ \frac{1}{4} (t+1)^2 \cos
^2\left(\frac{\pi  t}{2 t+2}\right) \Big], \\
u(\lambda,\alpha) &= \frac{2}{\pi} \int_0^{\infty}dt\,
e^{-\frac{t}{\lambda }}
\Big(\frac{\alpha}{\lambda}-\frac{(\alpha\lambda-1)}{\lambda}\partial_t\Big)
\Big[ \frac{1}{4} (t+1) \sin
\left(\frac{\pi  t}{t+1}\right) \Big], \\
v(\lambda,\alpha) &= \frac{2}{\pi} \int_0^{\infty}dt\,
e^{-\frac{t}{\lambda }}
\Big(\frac{\alpha}{\lambda}-\frac{(\alpha\lambda-1)}{\lambda}\partial_t\Big)
\Big[ \sin ^2\left(\frac{\pi  t}{2
t+2}\right) \Big], \\
w(\lambda,\alpha) &= \frac{2}{\pi} \int_0^{\infty}dt\,
e^{-\frac{t}{\lambda }}
\Big(\frac{\alpha}{\lambda}-\frac{(\alpha\lambda-1)}{\lambda}\partial_t\Big)
\Big[ \frac{1}{12} \left(3-t^2-2 t\right) \cos ^2\left(\frac{\pi
t}{2 t+2}\right) \Big].
\end{align}
These integrals are convergent provided that the parameter
$\lambda$ belongs to the interval $(0,+\infty)$, with $\alpha$
being an arbitrary parameter.

The numerical evaluation of the vacuum energy follows the same
steps developed in this subsection; in other words, we first
define the normalized value of the vacuum energy
\begin{eqnarray}
\label{normaener1zzextra} E_{\lambda,\alpha}(z) \equiv
\frac{\pi^2}{3} \langle z^{L_0} \Psi^{(1)}_{\lambda,\alpha}, Q_B
z^{L_0} \Psi^{(1)}_{\lambda,\alpha}\rangle,
\end{eqnarray}
we then plug (\ref{psilaal}) into (\ref{normaener1zzextra}), and
truncate the series to order $z^{2n-2}$, and, finally, we employ
Pad\'{e} approximants of order $P^{n}_{2+n}(\lambda,\alpha,z)$. We
have performed the numerical computations at level $n=8$ with
$\lambda=1/2$, $\lambda=1$ and for various values of the parameter
$\alpha$. The results are shown in table \ref{resultsx1extra}. We
see that the normalized value of the vacuum energy computed
numerically using Pad\'{e} approximants agrees with the answer
expected from Sen's first conjecture.
\begin{table}[ht]
\caption{The $P_{2+8}^{8}$ Pad\'{e} approximation for the
normalized value of the vacuum energy $\frac{\pi^2}{3} \langle
z^{L_0} \Psi^{(1)}_{\lambda,\alpha}, Q_B z^{L_0}
\Psi^{(1)}_{\lambda,\alpha}\rangle$ evaluated at various values of
the parameter $\alpha$, with $\lambda=1/2$ and $\lambda=1$.}
\centering \label{resultsx1extra}

\begin{tabular}{|c|c|c|}
\hline
  &  $P^{8}_{2+8}(\lambda = 1/2,\alpha,z=1)$  &  $P^{8}_{2+8}(\lambda = 1,\alpha,z=1)$  \\

\hline $\alpha=-1$   & $-1.003161547$ &  $-0.728764899$ \\
\hline $\alpha=0$  & $-0.994968610$ & $-0.795944823$ \\
\hline $\alpha=1$ & $-0.991643419$ & $-0.935655316$ \\
\hline $\alpha=1.5$   &  $-0.996141185$ & $-0.948563810$\\
\hline $\alpha=2$ & $-0.972530090$ & $-0.940860097$ \\
\hline
\end{tabular}

\end{table}

\section{Real one-parameter family of
solutions} In this section, we will study another one-parameter
family of solutions which satisfies the string field reality
condition. By performing a gauge transformation, this real
solution will be related to the previous non-real solution. The
level truncation computation of the vacuum energy will be analyzed
in a similar manner as for the non-real solution, namely by means
of Pad\'{e} approximants.

\subsection{Generating the real solution}
A solution $\hat \Psi_\lambda$ which satisfies the string field
reality condition can be generated by performing a gauge
transformation on the non-real solution (\ref{regularsol2})
\begin{eqnarray}
\label{gaugeyy1} \hat \Psi_\lambda =
\sqrt{\frac{1+(\lambda-1)K}{1+\lambda K}} (\Psi_\lambda + Q_B)
\sqrt{\frac{1+\lambda K}{1+(\lambda-1)K}}.
\end{eqnarray}
Because of the square root in (\ref{gaugeyy1}), we should be
careful about whether the solution derived from this equation is a
well-behaved element in the space of string fields introduced in
\cite{Schnabl:2010tb}. A standard criterion for determining
whether or not a string field is a well behaved element, consists
in studying the convergence properties of its level expansion
using the Virasoro basis of $L_0$ eigenstates.

It turns out that the real solution derived from (\ref{gaugeyy1})
\begin{eqnarray}
\label{gaugeyy2} \hat \Psi_\lambda =
\sqrt{\frac{1+(\lambda-1)K}{1+\lambda K}} (c+\lambda cKBc)
\sqrt{\frac{1+(\lambda-1)K}{1+\lambda K}},
\end{eqnarray}
as in the case of the non-real solution, can be written as an
expression containing an exact BRST term
\begin{align}
\label{exactbrstreal1} \hat\Psi_\lambda
=\sqrt{\frac{1+(\lambda-1)K}{1+\lambda K}} \,c\,
\sqrt{\frac{1+(\lambda-1)K}{1+\lambda K}}+ Q_B
\Big\{\sqrt{\frac{1+(\lambda-1)K}{1+\lambda K}} \lambda Bc
\sqrt{\frac{1+(\lambda-1)K}{1+\lambda K}} \Big\}.
\end{align}
Since we assume the validity of the string field equation of
motion when contracted with the solution itself, the exact BRST
term will not contribute to the evaluation of the vacuum energy.
So in the next subsection, we will only concentrate in the level
expansion of the first term of the real solution
(\ref{exactbrstreal1})
\begin{align}
\label{exactbrstreal2} \hat\Psi^{(1)}_\lambda
=\sqrt{\frac{1+(\lambda-1)K}{1+\lambda K}} \,c\,
\sqrt{\frac{1+(\lambda-1)K}{1+\lambda K}}.
\end{align}

\subsection{Level expansion of the real solution}
To expand the string field $\hat\Psi^{(1)}_\lambda$ in the
Virasoro basis of $L_0$ eigenstates, it is convenient to write an
integral representation for the square root function
\begin{eqnarray}
\label{rootsquare01} \sqrt{\frac{1+(\lambda-1)K}{1+\lambda K}} =
\int_0^{\infty}dt\, \frac{e^{\frac{t (1-2 \lambda )}{2 (\lambda
-1) \lambda }}
   I_0\left(\frac{1}{2 (\lambda -1) \lambda } t \right)}{\sqrt{(\lambda -1)
   \lambda }} \big[1-(\lambda-1)\partial_t\big]e^{-K t} \,,
\end{eqnarray}
where $I_0(x)$ is the modified Bessel function of the first kind.
This integral is well defined if the parameter $\lambda
>1$. Since the limit case $\lambda \rightarrow 1$ corresponds to
the function $1/\sqrt{1+K}$, we should have
\cite{Erler:2009uj}\footnote{In fact this limit can be derived by
using the asymptotic expansion $I_0(x)\approx e^x/\sqrt{2 \pi
x}\big[1+O(1/x)\big]$ which is valid for $x\gg 1$.}
\begin{eqnarray}
\label{rootsquare02} \lim_{\lambda \rightarrow 1}
\Big[\frac{e^{\frac{t (1-2 \lambda )}{2 (\lambda -1)
   \lambda }} I_0\left(\frac{1}{2 (\lambda -1)
   \lambda }t\right)}{\sqrt{(\lambda -1) \lambda }}\Big] =
\frac{e^{-t}}{\sqrt{\pi t}}.
\end{eqnarray}

Plugging the definition of the square root function
(\ref{rootsquare01}) into the expression for the string field
(\ref{exactbrstreal2}), we get
\begin{align}
\label{rootsquarelevel01} \hat\Psi^{(1)}_\lambda =
\int_0^{\infty}dsdt\, \frac{e^{-\frac{(s+t) (2 \lambda -1)}{2
(\lambda -1)
   \lambda }} I_0\left(\frac{1}{2 (\lambda -1)
   \lambda }s\right) I_0\left(\frac{1}{2 (\lambda -1)
   \lambda }t\right)}{(\lambda -1) \lambda }\big[1-(\lambda-1)\partial_s\big]\big[1-(\lambda-1)\partial_t\big]
 \Omega^s c \Omega^t \,. \;\;\;\;
\end{align}
Using equation (\ref{c01}), the representation for the wedge state
(\ref{omeome1}), and noting that under the conformal map $\tilde
z=\frac{2}{\pi} \arctan z$ the $c$ ghost transforms as $\tilde
c(\tilde z)=\frac{2}{\pi} \cos^{2}(\frac{\pi}{2} \tilde z) c
\big(\tan (\frac{\pi}{2} \tilde z)\big)$ from equation
(\ref{rootsquarelevel01}), we obtain
\begin{align}
\label{psilevelrootxx2} \hat\Psi^{(1)}_\lambda =\frac{2}{\pi}
\int_0^{\infty}dsdt\,
\mathcal{I}(s,t,\lambda)\widehat{\mathcal{O}}_s
\widehat{\mathcal{O}}_t \Big[\mathcal{J}(s,t)\,
\widetilde{U}_{s+t+1}\, c \Big(\frac{2}{s+t+1} \tan
\big(\frac{\pi}{2}\frac{ s-t}{s+t+1}\big) \Big) \Big]| 0 \rangle ,
\end{align}
where the functions $\mathcal{I}(s,t,\lambda)$, $\mathcal{J}(s,t)$
and the operators $\widehat{\mathcal{O}}_t$,
$\widetilde{U}_{s+t+1}$, are defined as follows
\begin{eqnarray}
\label{aux001} \mathcal{I}(s,t,\lambda)&=&\frac{e^{-\frac{(s+t) (2
\lambda -1)}{2 (\lambda -1)
   \lambda }} I_0\left(\frac{1}{2 (\lambda -1)
   \lambda }s\right) I_0\left(\frac{1}{2 (\lambda -1)
   \lambda }t\right)}{(\lambda -1) \lambda }, \\
\label{aux002} \mathcal{J}(s,t)&=&\big(\frac{s+t+1}{2}\big)^2 \cos
^2\left(\frac{\pi}{2}\frac{ s-t}{s+t+1}\right), \\
\label{aux003} \widetilde{U}_{s+t+1} &=& \cdots e^{u_{10,s+t+1}
L_{-10}} e^{u_{8,s+t+1} L_{-8}} e^{u_{6,s+t+1} L_{-6}}
e^{u_{4,s+t+1}
L_{-4}}e^{u_{2,s+t+1} L_{-2}}, \\
\label{aux004}\widehat{\mathcal{O}}_t
&=&\big[1-(\lambda-1)\partial_t\big].
\end{eqnarray}

By writing the $c$ ghost in terms of its modes $c(z)=\sum_{m}
c_m/z^{m-1}$ and employing equations
(\ref{psilevelrootxx2})-(\ref{aux004}), we can expand
$\hat\Psi^{(1)}_\lambda$ in terms of elements contained in the
Virasoro basis of $L_0$ eigenstates. For instance, let us expand
$\hat\Psi^{(1)}_\lambda$ up to level two states
\begin{eqnarray}
\hat\Psi^{(1)}_\lambda= t(\lambda) c_1 | 0 \rangle + v(\lambda)
c_{-1} | 0 \rangle + w(\lambda) L_{-2}c_1 | 0 \rangle + \cdots ,
\end{eqnarray}
where the coefficients of the expansion $t(\lambda)$, $v(\lambda)$
and $w(\lambda)$  are given by the following integrals
\begin{align}
t(\lambda) &= \frac{2}{\pi} \int_0^{\infty}dsdt\,
\mathcal{I}(s,t,\lambda)\widehat{\mathcal{O}}_s
\widehat{\mathcal{O}}_t \Big[\mathcal{J}(s,t)\Big], \\
v(\lambda) &= \frac{2}{\pi} \int_0^{\infty}dsdt\,
\mathcal{I}(s,t,\lambda)\widehat{\mathcal{O}}_s
\widehat{\mathcal{O}}_t \Big[\frac{4\mathcal{ J}(s,t) \tan ^2\left(\frac{\pi}{2}\frac{s-t}{s+t+1}\right)}{(s+t+1)^2}\Big], \\
w(\lambda) &= \frac{2}{\pi} \int_0^{\infty}dsdt\,
\mathcal{I}(s,t,\lambda)\widehat{\mathcal{O}}_s
\widehat{\mathcal{O}}_t \Big[\frac{\big(4-(s+t+1)^2\big)
\mathcal{J}(s,t)}{3 (s+t+1)^2}\Big].
\end{align}
These integrals are convergent provided that the parameter
$\lambda$ belongs to the interval $(1,+\infty)$. Employing
numerical values for the parameter $\lambda$, the integrals can be
evaluated numerically. The limit case $\lambda \rightarrow 1$
precisely corresponds to the Erler-Schnabl's twist even (real)
solution which was analyzed in \cite{Erler:2009uj}.

At this point we would like to explain why the symmetric solution
does not exist for $\lambda < 1$. The answer is related to the
holomorphicity property of the square root function $F(K)$, which
is defined on the left-hand side of (\ref{rootsquare01}). Recall
that a function $f$ defined on a non-empty open set $\mathcal{O}$
is holomorphic if its derivative $f'$ is well defined at every
point in $\mathcal{O}$. For our case, the derivative
\begin{eqnarray}
\label{functionderi01} F'(K)=-\frac{1}{2 \big(1+(\lambda
-1)K\big)^{1/2} \big(1+ \lambda K\big)^{3/2}}
\end{eqnarray}
is well defined in the region where $\lambda > 1$ and
$\mathrm{Re}\,K>0$, whereas in the region where $\lambda <1$ and
$\mathrm{Re}\,K>0$ it is not well defined. Therefore the left-hand
side of (\ref{rootsquare01}) is not holomorphic in the region
where $\lambda <1$ and $\mathrm{Re}\,K>0$, and consequently using
the criterion for safe string fields derived in reference
\cite{Schnabl:2010tb}, the symmetric solution does not exist.

\subsection{Evaluation of the vacuum energy}
In this subsection, in order to evaluate the normalized value of
the vacuum energy for the real solution
\begin{eqnarray}
\label{normaenerreal1} E = \frac{\pi^2}{3} \langle
\hat{\Psi}^{(1)}_\lambda, Q_B \hat{\Psi}^{(1)}_\lambda\rangle,
\end{eqnarray}
we will perform the replacement $\hat{\Psi}^{(1)}_\lambda
\rightarrow z^{L_0} \hat{\Psi}^{(1)}_\lambda$ in the $L_0$ level
truncation scheme, so that states in the $L_0$ level expansion of
the solution will acquire different integer powers of $z$ at
different levels. The parameter $z$ is needed because we will need
to express the normalized value of the vacuum energy
(\ref{normaenerreal1}) as a formal power series expansion if we
want to use Pad\'{e} approximants
\cite{Arroyo:2009ec,AldoArroyo:2008zm}. After doing our
calculations, we will simply set $z = 1$.

Using the equations (\ref{psilevelrootxx2})-(\ref{aux004}), the
string field $\hat{\Psi}^{(1)}_\lambda$ can be readily expanded
and the individual coefficients can be numerically integrated. For
instance, employing some numerical values for the parameter
$\lambda$, we obtain
\begin{align}
\label{psireal6over5} \hat{\Psi}^{(1)}_{\lambda=6/5} &=&
0.48079059\, c_1| 0 \rangle+0.15487181\, c_{-1}| 0
\rangle+0.00031916\, L_{-2}c_1| 0 \rangle + 0.13495200\, c_{-3}| 0 \rangle \nonumber \\
&& -0.01975863\, L_{-4}c_1| 0 \rangle-0.00920921\,L_{-2}c_{-1}| 0
\rangle+0.03282467\,L_{-2}L_{-2}c_1| 0 \rangle+\cdots,\\
\label{psireal7over5} \hat{\Psi}^{(1)}_{\lambda=7/5} &=&
0.45869460\, c_1| 0 \rangle+0.16142345\, c_{-1}| 0
\rangle+0.00550055\, L_{-2}c_1| 0 \rangle + 0.15824508\, c_{-3}| 0 \rangle \nonumber \\
&& -0.02526528\, L_{-4}c_1| 0 \rangle-0.01365061\,L_{-2}c_{-1}| 0
\rangle+0.04027528\,L_{-2}L_{-2}c_1| 0 \rangle+\cdots,\\
\label{psirreal8over5} \hat{\Psi}^{(1)}_{\lambda=8/5} &=&
0.44104558\, c_1| 0 \rangle+0.16132987\, c_{-1}| 0
\rangle+0.01141476\, L_{-2}c_1| 0 \rangle + 0.17029934\, c_{-3}| 0 \rangle \nonumber \\
&& -0.02990375\, L_{-4}c_1| 0 \rangle-0.01729818\,L_{-2}c_{-1}| 0
\rangle+0.04603466\,L_{-2}L_{-2}c_1| 0 \rangle+\cdots,\\
\label{psireal9over5} \hat{\Psi}^{(1)}_{\lambda=9/5} &=&
0.42662772\, c_1| 0 \rangle+0.15821733\, c_{-1}| 0
\rangle+0.01725819\, L_{-2}c_1| 0 \rangle + 0.17597281\, c_{-3}| 0 \rangle \nonumber \\
&& -0.03380422\, L_{-4}c_1| 0 \rangle-0.01997450\,L_{-2}c_{-1}| 0
\rangle+0.05058763\,L_{-2}L_{-2}c_1| 0 \rangle+\cdots,\\
\label{psireal2} \hat{\Psi}^{(1)}_{\lambda=2} &=& 0.41461941\,
c_1| 0 \rangle+0.15372018\, c_{-1}| 0
\rangle+0.02275988\, L_{-2}c_1| 0 \rangle + 0.17784600\, c_{-3}| 0 \rangle \nonumber \\
&& -0.03711080\, L_{-4}c_1| 0 \rangle-0.02185865\,L_{-2}c_{-1}| 0
\rangle+0.05426470\,L_{-2}L_{-2}c_1| 0 \rangle+\cdots.
\end{align}

Once we have the level expansion of the string field
$\hat{\Psi}^{(1)}_\lambda$, we can compute the normalized value of
the vacuum energy. By performing the replacement
$\hat{\Psi}^{(1)}_\lambda \rightarrow z^{L_0}
\hat{\Psi}^{(1)}_\lambda$ from equation (\ref{normaenerreal1}), we
define
\begin{eqnarray}
\label{normaenerreal1zz} E_\lambda(z) \equiv \frac{\pi^2}{3}
\langle z^{L_0} \hat{\Psi}^{(1)}_\lambda, Q_B  z^{L_0}
\hat{\Psi}^{(1)}_\lambda\rangle.
\end{eqnarray}
For example, plugging the level expansions
(\ref{psireal6over5})-(\ref{psireal2}) into the definition
(\ref{normaenerreal1zz}), we obtain
\begin{eqnarray}
\label{normaener1zzreal6over5} E_{\lambda=6/5}(z)
&=&-\frac{0.76048459}{z^2}-0.07793135 z^2-0.25756015
z^6+\cdots, \;\;\;\;\; \\
\label{normaener1zzreal7over5} E_{\lambda=7/5}(z)
&=&-\frac{0.69219068}{z^2}-0.06780088 z^2-0.38256079
z^6+\cdots, \;\;\;\;\; \\
\label{normaener1zzreal8over5} E_{\lambda=8/5}(z)
&=&-\frac{0.63994910}{z^2}-0.04756125 z^2-0.48705471
z^6+\cdots, \;\;\;\;\; \\
\label{normaener1zzreal9over5} E_{\lambda=9/5}(z)
&=&-\frac{0.59879290}{z^2}-0.02453608 z^2-0.57161143
z^6+\cdots, \;\;\;\;\;\\
\label{normaener1zzreal2} E_{\lambda=2}(z)
&=&-\frac{0.56555878}{z^2}-0.00186180 z^2-0.63933036 z^6+\cdots.
\;\;\;\;\;
\end{eqnarray}

Using these kind of series in $z$ for $E_{\lambda}(z)$, by the
standard procedure based on Pad\'{e} approximants of order
$P^n_{2+n}(\lambda,z)$, we numerically compute the normalized
value of the vacuum energy. The label $n$ corresponds to the power
of $z$ in the series expansion of $E_{\lambda}(z)$ truncated up to
the order $z^{2n-2}$. The results of our calculations are
summarized in table \ref{realresultsx1}. As we can see, the
normalized value of the vacuum energy evaluated using Pad\'{e}
approximants nicely confirm Sen's first conjecture
\cite{Sen:1999mh,Sen:1999nx}.

\begin{table}[ht]
\caption{The Pad\'{e} approximation for the normalized value of
the vacuum energy $\frac{\pi^2}{3} \langle z^{L_0}
\hat{\Psi}^{(1)}_\lambda, Q_B z^{L_0}
\hat{\Psi}^{(1)}_\lambda\rangle$ evaluated at $z=1$. The results
show the $P_{2+n}^{n}$ Pad\'{e} approximation for various values
of the parameter $\lambda$.} \centering
\begin{tabular}{|c|c|c|c|c|c|}
\hline
  &  $P^{n}_{2+n}[\lambda = \frac{11}{10}]$   &  $P^{n}_{2+n}[\lambda = \frac{12}{10}]$ &
  $P^{n}_{2+n}[\lambda = \frac{13}{10} ]$ & $P^{n}_{2+n}[\lambda = \frac{14}{10} ]$ & $P^{n}_{2+n}[\lambda = \frac{15}{10}]$ \\
    \hline
$n=0$   &  $-0.80293844$ &  $-0.76048459$ & $-0.72392265$ & $-0.69219068$ & $-0.66442720$\\
\hline  $n=4$ & $-0.75437392$  & $-0.72667434$ & $-0.70128551$ & $-0.67758602$ & $-0.65544662$  \\
\hline $n=8$   & $-0.98896263$ & $-0.98497033$ & $-0.98008354$ & $-0.97427446$ & $-0.96760968$ \\
\hline $n=12$ & $-0.99001621$ & $-0.98660724$ & $-0.98251492$ & $-0.97749570$ & $-0.97151457$ \\
\hline
\end{tabular}
\label{realresultsx1}

\begin{tabular}{|c|c|c|c|c|c|}
\hline
  &  $P^{n}_{2+n}[\lambda = \frac{16}{10}]$   &  $P^{n}_{2+n}[\lambda = \frac{17}{10}]$ &
  $P^{n}_{2+n}[\lambda = \frac{18}{10} ]$ & $P^{n}_{2+n}[\lambda = \frac{19}{10} ]$ & $P^{n}_{2+n}[\lambda = 2 ]$ \\
    \hline
$n=0$   &  $-0.63994910$ &  $-0.61821502$ & $-0.59879290$ & $-0.58133505$ & $-0.56555878$\\
\hline  $n=4$ & $-0.63480210$  & $-0.61557936$ & $-0.59769248$ & $-0.58104857$ & $-0.56555334$  \\
\hline $n=8$   & $-0.96019541$ & $-0.95214843$ & $-0.94358184$ & $-0.93459932$ & $-0.92529280$ \\
\hline $n=12$ & $-0.96462560$ & $-0.95692661$ & $-0.94853355$ & $-0.93956627$ & $-0.93013994$ \\
\hline
\end{tabular}
\label{realresultsx2}
\end{table}

\subsection{Closed string tadpole}
There is another gauge invariant quantity which is known in the
literature as the closed string tadpole
\cite{Ellwood:2008jh,Hashimoto:2001sm}. This quantity was already
evaluated in \cite{Zeze:2010sr} for the case of the non-real
solution (\ref{exactbrst1}), and it was shown that its value
coincides with the expected answer of closed string tadpole on the
disk \cite{Ellwood:2008jh}.

In this subsection, we would like to perform a similar computation
for the case of the real solution
\begin{align}
\label{exactbrstreal1closed} \hat\Psi_\lambda
=\sqrt{\frac{1+(\lambda-1)K}{1+\lambda K}} \,c\,
\sqrt{\frac{1+(\lambda-1)K}{1+\lambda K}}+ Q_B
\Big\{\sqrt{\frac{1+(\lambda-1)K}{1+\lambda K}} \lambda Bc
\sqrt{\frac{1+(\lambda-1)K}{1+\lambda K}} \Big\}.
\end{align}

The second term on the right hand side of
(\ref{exactbrstreal1closed}) does not contribute to the tadpole
due to the BRST invariance of the tadpole. Then we are going to
compute
\begin{eqnarray}
\label{tadpole01} \text{Tr}[V\hat\Psi_\lambda]= \text{Tr}\Big[ V
\sqrt{\frac{1+(\lambda-1)K}{1+\lambda K}} \,c\,
\sqrt{\frac{1+(\lambda-1)K}{1+\lambda K}}\Big],
\end{eqnarray}
where $V=c \tilde c \mathcal{V}_{\text{matter}}$ is a closed
string vertex operator insertion at the open string midpoint. We
can write an integral representation for the square root function
\begin{eqnarray}
\label{rootsquare01zz} \sqrt{\frac{1+(\lambda-1)K}{1+\lambda K}} =
\int_0^{\infty}dt\, \frac{e^{\frac{t (1-2 \lambda )}{2 (\lambda
-1) \lambda }}
   I_0\left(\frac{1}{2 (\lambda -1) \lambda } t \right)}{\sqrt{(\lambda -1)
   \lambda }} \big[1-(\lambda-1)\partial_t\big]e^{-K t} \,,
\end{eqnarray}
where $I_0(x)$ is the modified Bessel function of the first kind.
Using (\ref{rootsquare01zz}), from equation (\ref{tadpole01}) we
get
\begin{align}
\label{rootsquarelevel01zz} \text{Tr}[V\hat\Psi_\lambda] =
\int_0^{\infty}dsdt\, \mathcal{I}(s,t,\lambda)
\big[1-(\lambda-1)\partial_s\big]\big[1-(\lambda-1)\partial_t\big]
 \text{Tr}[V\Omega^s c \Omega^t] \,,
\end{align}
where the function $\mathcal{I}(s,t,\lambda)$ was defined in
(\ref{aux001}).

The inner product $\text{Tr}[V\Omega^s c \Omega^t]$ is a
correlator on a cylinder of circumference $s+t$; by a scale
transformation we can reduce it to a cylinder of unit
circumference, producing a factor of $s+t$ for the c ghost from
the conformal transformation. Thus $\text{Tr}[V\hat\Psi_\lambda]$
can be written as
\begin{align}
\label{rootsquarelevel01x2}\text{Tr}[V\hat\Psi_\lambda] =
\int_0^{\infty}dsdt\,
\mathcal{I}(s,t,\lambda)\big[1-(\lambda-1)\partial_s\big]\big[1-(\lambda-1)\partial_t\big]\big((s+t)
 \text{Tr}[Vc \Omega]\big) \,.
\end{align}
From this equation (\ref{rootsquarelevel01x2}), we obtain
\begin{align}
\label{rootsquarelevel01x3} \text{Tr}[V\hat\Psi_\lambda]=
\text{Tr}[Vc \Omega]\times \int_0^{\infty}dsdt\,
\frac{e^{-\frac{(s+t) (2 \lambda -1)}{2 (\lambda -1) \lambda }}
(s+t-2 \lambda +2)
   I_0\left(\frac{s}{2 (\lambda -1) \lambda }\right) I_0\left(\frac{t}{2 (\lambda -1) \lambda
   }\right)}{(\lambda -1) \lambda }.
\end{align}
This double integral is well defined for $\lambda > 1$ and it can
be analytically computed giving as a result a value which does not
depend on $\lambda$
\begin{align}
\label{rootsquarelevel01x4}\int_0^{\infty}dsdt\,
\frac{e^{-\frac{(s+t) (2 \lambda -1)}{2 (\lambda -1) \lambda }}
(s+t-2 \lambda +2)
   I_0\left(\frac{s}{2 (\lambda -1) \lambda }\right) I_0\left(\frac{t}{2 (\lambda -1) \lambda
   }\right)}{(\lambda -1) \lambda } = 1.
\end{align}
Plugging the answer of this integral (\ref{rootsquarelevel01x4})
into the equation (\ref{rootsquarelevel01x3}), we get the desired
result
\begin{align}
\label{rootsquarelevel01x5} \text{Tr}[V\hat\Psi_\lambda]=
\text{Tr}[Vc \Omega] = \langle
\mathcal{V}(i\infty)c(0)\rangle_{C_1}.
\end{align}
This result, as similar to the case of the non-real solution
\cite{Zeze:2010sr}, coincides with the expected answer of closed
string tadpole on the disk \cite{Ellwood:2008jh}.

\section{Summary and discussion}

Using states in the Virasoro basis of $L_0$ eigenstates, we have
analyzed the level expansion of regularized identity based
solutions in open bosonic string field theory. We have evaluated
the vacuum energy for these solutions by means of the numerical
method based on Pad\'{e} approximants. Our results confirm the
expected answer for the tachyon vacuum in agreement with Sen's
first conjecture.

In order to simplify the computation of the vacuum energy, we have
assumed the validity of the string field equations of motion when
contracted with the regularized solution itself. This assumption
has been proven correct by explicit evaluation of correlation
functions in the so-called $KBc$ subalgebra and by using
$\mathcal{L}_0$ level truncation computations
\cite{Arroyo:2010sy}. Nevertheless, it should be nice to confirm
the validity of this assumption by using a third option, namely by
employing the usual Virasoro $L_0$ level truncation scheme.

The procedures developed in this work, related to the level
truncation analysis of regularized identity based solutions, can
be extended for solutions in the case of the modified cubic
superstring field theory \cite{Arefeva:1989cp}, as well as in
Berkovits non-polynomial open superstring field theory
\cite{Noumi:2011kn,Berkovits:1995ab}. Since in the case of the
modified cubic superfield theory, we have already proposed a way
for regularizing identity based solutions \cite{Arroyo:2010sy},
the level truncation analysis should be straightforward. However,
constructing a similar regularization for Berkovits superstring
field theory remains an unsolved problem.

\section*{Acknowledgements}
I would like to thank Ted Erler, Michael Kiermaier, Isao Kishimoto
and Michael Kroyter for useful discussions. I also thank the
Institute of Physics of the Academy of Sciences of Czech Republic
for hospitality during the International Conference on String
Field Theory and Related Aspects SFT 2011, where this work was
completed. Finally I would like to give a special thank to my wife
Diany and my newborn son Davi for their valuable company during
the elaboration of this work. This work is supported by FAPESP
grant 2010/05990-0.



\begin{thebibliography}
\bibitem{Arroyo:2010sy}
  E.~A.~Arroyo,\textit{ Comments on regularization of identity based solutions in string
  field theory}, JHEP 1011, 135 (2010), [arXiv:1009.0198].

\bibitem{Zeze:2010sr}
  S.~Zeze, \textit{Regularization of identity based solution in string field
theory}, JHEP 1010, 070 (2010), [arXiv:1008.1104].

\bibitem{Witten:1985cc}
E. Witten, \textit{Noncommutative Geometry And String Field
Theory}, Nucl. Phys. B 268 (1986) 253.

\bibitem{Arroyo:2010fq}
  E.~A.~Arroyo, \textit{Generating Erler-Schnabl-type Solution for Tachyon Vacuum in Cubic
  Superstring Field Theory},
  J.\ Phys.\ A  43, 445403 (2010), [arXiv:1004.3030].

\bibitem{Erler:2009uj}
  T. Erler and M. Schnabl, \textit{A Simple Analytic Solution for Tachyon Condensation},
  JHEP 0910, 066 (2009),
  [arXiv:0906.0979].

\bibitem{Zeze:2011zz}
  S.~Zeze, \textit{Application of KBc subalgebra in string field theory},
  Prog.\ Theor.\ Phys.\ Suppl.\  188, 56 (2011).

\bibitem{Kostelecky:1988ta}
V. A. Kostelecky and S. Samuel, \textit{The Static Tachyon
Potential In The Open Bosonic String Theory}, Phys. Lett. B 207,
169 (1988).

\bibitem{Kostelecky:1989nt}
V. A. Kostelecky and S. Samuel, \textit{On A Nonperturbative
Vacuum For The Open Bosonic String}, Nucl. Phys. B 336, 263
(1990).

\bibitem{Moeller:2000xv}
N. Moeller and W. Taylor, \textit{Level truncation and the tachyon
in open bosonic string field theory}, Nucl. Phys. B 583, 105
(2000), hep-th/0002237.

\bibitem{Taylor:2002fy}
  W.~Taylor,
  \textit{A perturbative analysis of tachyon condensation},
  JHEP 0303, 029 (2003), hep-th/0208149.

\bibitem{Gaiotto:2002wy}
D. Gaiotto and L. Rastelli, \textit{Experimental string field
theory}, JHEP 0308, 048 (2003), hep-th/0211012.

\bibitem{Kishimoto:2011zz}
  I.~Kishimoto, \textit{On numerical solutions in open string field theory},
  Prog.\ Theor.\ Phys.\ Suppl.\  188 (2011) 155.

\bibitem{Murata:2011ex}
  M.~Murata and M.~Schnabl, \textit{On Multibrane Solutions in Open String Field Theory},
  Prog.\ Theor.\ Phys.\ Suppl.\  188, 50 (2011), [arXiv:1103.1382].

\bibitem{Schnabl:2010tb}
  M.~Schnabl, \textit{Algebraic solutions in Open String Field Theory - a lightning review}, [arXiv:1004.4858].

\bibitem{Takahashi:2007du}
  T.~Takahashi, \textit{Level truncation analysis of exact solutions in open string field
theory},
  JHEP 0801, 001 (2008)
  [arXiv:0710.5358].

\bibitem{Schnabl:2002gg}
  M.~Schnabl, \textit{Wedge states in string field theory},
  JHEP 0301, 004 (2003), hep-th/0201095.

\bibitem{Schnabl:2005gv}
  M.~Schnabl,
 \textit{ Analytic solution for tachyon condensation in open string field theory},
  Adv.\ Theor.\ Math.\ Phys.\  10, 433 (2006), hep-th/0511286.

\bibitem{Kiermaier:2007ba}
  M.~Kiermaier, Y.~Okawa, L.~Rastelli and B.~Zwiebach, \textit{Analytic solutions for marginal deformations in open string field
theory}, JHEP 0801, 028 (2008), hep-th/0701249.

\bibitem{Erler:2011tc}
  T.~Erler and C.~Maccaferri, \textit{Comments on Lumps from RG flows}, [arXiv:1105.6057].

\bibitem{Bonora:2011ru}
  L.~Bonora, S.~Giaccari and D.~D.~Tolla, \textit{Analytic solutions for Dp branes in SFT},
  [arXiv:1106.3914].

\bibitem{Bonora:2011ri}
  L.~Bonora, S.~Giaccari and D.~D.~Tolla, \textit{The energy of the analytic lump solution in SFT},
  [arXiv:1105.5926].

\bibitem{Okawa:2006vm}
Y. Okawa, \textit{Comments on Schnabl's analytic solution for
tachyon condensation in Witten's open string field theory}, JHEP
0604, 055 (2006), hep-th/0603159.

\bibitem{Fuchs:2006hw}
E. Fuchs and M. Kroyter, \textit{On the validity of the solution
of string field theory}, JHEP 0605, 006 (2006), hep-th/0603195.

\bibitem{Arroyo:2009ec}
  E. A. Arroyo, \textit{Cubic interaction term for Schnabl's solution using Pade
approximants}, J.\ Phys.\ A  42, 375402 (2009),
  [arXiv:0905.2014].

\bibitem{Erler:2006hw}
  T.~Erler, \textit{Split string formalism and the closed string vacuum},
  JHEP 0705, 083 (2007), hep-th/0611200.

\bibitem{Erler:2006ww}
  T. Erler, \textit{Split string formalism and the closed string vacuum. II},
  JHEP 0705, 084 (2007), hep-th/0612050.

\bibitem{Arroyo:2011zt}
  E.~A.~Arroyo, \textit{Conservation laws and tachyon potentials in the sliver frame}, JHEP 1106, 033
  (2011), [arXiv:1103.4830].

\bibitem{AldoArroyo:2009hf}
  E.~Aldo Arroyo, \textit{The Tachyon Potential in the Sliver Frame}, JHEP 0910, 056 (2009), [arXiv:0907.4939].

\bibitem{Sen:1999mh}
  A.~Sen, \textit{Descent relations among bosonic D-branes}, Int.\ J.\ Mod.\ Phys.\  A 14, 4061
  (1999), hep-th/9902105.

\bibitem{Sen:1999nx}
A. Sen and B. Zwiebach, \textit{Tachyon condensation in string
field theory}, JHEP 0003, 002 (2000), hep-th/9912249.

\bibitem{Bonora:2011ns}
  L.~Bonora, S.~Giaccari and D.~D.~Tolla, \textit{Lump solutions in SFT. Complements},
  [arXiv:1109.4336].

\bibitem{Gaberdiel:1997ia}
  M.~R.~Gaberdiel and B.~Zwiebach, \textit{Tensor constructions of open string theories. 1: Foundations},
  Nucl.\ Phys.\  B 505, 569 (1997), hep-th/9705038.

\bibitem{AldoArroyo:2008zm}
  E.~Aldo Arroyo, \textit{Pure Spinor Partition Function Using Pade Approximants}, JHEP 0807, 081 (2008), [arXiv:0806.0643].

\bibitem{Ellwood:2008jh}
  I.~Ellwood, \textit{The Closed string tadpole in open string field theory},
  JHEP 0808, 063 (2008), [arXiv:0804.1131].

\bibitem{Hashimoto:2001sm}
  A.~Hashimoto and N.~Itzhaki, \textit{Observables of string field theory},
  JHEP 0201, 028 (2002), hep-th/0111092.

\bibitem{Arefeva:1989cp}
  I.~Y.~Arefeva, P.~B.~Medvedev and A.~P.~Zubarev, \textit{New Representation For String Field Solves The Consistency Problem For Open Superstring Field Theory},
  Nucl.\ Phys.\  B 341, 464 (1990).

\bibitem{Noumi:2011kn}
  T.~Noumi and Y.~Okawa, \textit{Solutions from boundary condition changing operators in open
  superstring field theory}, [arXiv:1108.5317].

\bibitem{Berkovits:1995ab}
  N.~Berkovits, \textit{SuperPoincare invariant superstring field theory},
  Nucl.\ Phys.\  B 450, 90 (1995),
  [Erratum-ibid.\  B 459, 439 (1996)],
  hep-th/9503099.

\end{thebibliography}
\end{document}